\documentclass[fleqn,usenatbib]{mnras}

\usepackage{newtxtext,newtxmath}

\usepackage[T1]{fontenc}
\usepackage{ae,aecompl}

\usepackage{graphicx}	
\usepackage{amsmath}	

\title[AM CVn Accretion States]{Evidence that short period AM CVn systems are diverse in outburst behaviour}

\author[C. Duffy et al.]{C. Duffy,$^{1,2}$\thanks{Contact e-mail: \href{mailto:christopher.duffy@armagh.ac.uk}{christopher.duffy@armagh.ac.uk}}
G. Ramsay,$^{1}$
D. Steeghs,$^{2}$
V. Dhillon,$^{3, 4}$
M. R. Kennedy,$^{5}$
D. Mata S\'anchez,$^{5}$
\newauthor
K. Ackley,$^{6,7}$
M. Dyer,$^{3}$
J. Lyman,$^{2}$
K. Ulaczyk,$^{2}$
D. K. Galloway,$^{6,7}$
P. O'Brien,$^{8}$
\newauthor
K. Noysena,$^{9}$
L. Nuttall,$^{10}$
D. Pollacco$^{2}$\\ \\ 
$^{1}$Armagh Observatory and Planetarium, College Hill, Armagh, BT61 9DB, UK\\
$^{2}$Department of Physics, University of Warwick, Gibbet Hill Road, Coventry, CV4 7AL, UK\\
$^{3}$Department of Physics and Astronomy, Hicks Building, The University of Sheffield, Sheffield, S3 7RH, UK\\
$^{4}$Instituto de Astrof\'{i}sica de Canarias, 38205 La Laguna, Tenerife, Spain\\
$^{5}$Jodrell Bank Centre for Astrophysics, Department of Physics and Astronomy, The University of Manchester, Manchester, M13 9PL, UK\\
$^{6}$School of Physics \& Astronomy, Monash University, Clayton VIC 3800, Australia\\
$^{7}$OzGrav: The ARC Centre of Excellence for Gravitational Wave Discovery, Clayton VIC 3800, Australia\\
$^{8}$School of Physics and Astronomy, University of Leicester, University Road, Leicester, LE1 7RH, UK\\
$^{9}$National Astronomical Research Institute of Thailand,  260  Moo 4, T. Donkaew,  A. Maerim, Chiangmai, 50180, Thailand\\
$^{10}$Institute of Cosmology and Gravitation, University of Portsmouth, Dennis Sciama Building, Burnaby Road, Portsmouth, PO1 3FX, UK}

\date{Accepted 2021 February 05. Received 2021 February 05; in original form 2020 December 18 }

\pubyear{2021}

\begin{document}
\label{firstpage}
\pagerange{\pageref{firstpage}--\pageref{lastpage}}
\maketitle

\begin{abstract}
We present results of our analysis of up to 15 years of photometric data from eight AM CVn systems with orbital periods between 22.5 and 26.8 min. Our data has been collected from the GOTO, ZTF, Pan-STARRS, ASAS-SN and Catalina all-sky surveys and amateur observations collated by the AAVSO. We find evidence that these interacting ultra-compact binaries show a similar diversity of long term optical properties as the hydrogen accreting dwarf novae. We found that AM CVn systems in the previously identified accretion disc instability region are not a homogenous group. Various members of the analysed sample exhibit behaviour reminiscent of Z Cam systems with long super outbursts and standstills, SU UMa systems with regular, shorter super outbursts, and nova-like systems which appear only in a high state. The addition of TESS full frame images of one of these systems, KL Dra, reveals the first evidence for normal outbursts appearing as a precursor to super outbursts in an AM CVn system. Our results will inform theoretical modelling of the outbursts of hydrogen deficient systems.
\end{abstract}

\begin{keywords}
accretion, accretion discs -- binaries: close -- stars: dwarf novae -- surveys
\end{keywords}



\section{Introduction}
AM CVn systems are at the shortest tail of the cataclysmic variable (CV) period distribution, consisting of a white dwarf which accretes matter from a hydrogen deficient low mass companion that is either fully or partially degenerate. The orbital periods of these systems are in the range $\sim5-65$ mins \citep{2010PASP..122.1133S}. The first AM CVn was identified in 1967 \citep{1967AcA....17..255S,1975AcA....25..371S} and by 2018 the known population consisted of 56 systems \citep{2018A&A...620A.141R}, with a few more in subsequent years. It is understood that the accretion is driven by gravitational wave (GW) radiation as the binary system loses angular momentum, as first proposed by \citet{1962ApJ...136..312K}. This GW radiation is predicted to be detectable by LISA \citep{2006CQGra..23S.809S} which intends to employ AM CVn systems as verification sources. The expected GW signal can be predicted from the distance, derived from Gaia parallax data, and the component masses, derived from photometric and spectroscopic observations, \citep{2018MNRAS.480..302K}. This utility in part drives the desire to improve our understanding of AM CVn systems.

In addition to these aforementioned properties understood to be common to all AM CVn systems, some systems have shown outbursting behaviour. These events are characterised by a sudden increase in brightness of 3-4 magnitudes which often leads to their discovery. It has previously been predicted and subsequently observed that systems with orbital periods between approximately 22 and 44 minutes exhibit outburst behaviour \citep{2012MNRAS.419.2836R}. This correlates and agrees well with predictions of the behaviour of the accretion discs of AM CVn systems, and how they vary with orbital period. Those systems with the shortest periods are expected to have hot, small and stable accretion discs and those with the longest periods are expected to have cool, large and stable accretion discs. Those systems with intermediate periods are expected, however, to have an unstable disc which can act as the source of outbursts \citep{2010PASP..122.1133S,2012A&A...544A..13K}.

Outbursting AM CVn systems share a number of characteristics with hydrogen dominated dwarf nova, whereby they both exhibit so-called normal and ``super outbursts'' (SO). SO last several weeks or more, e.g. $\sim10$ days in KL Dra or $\sim29$ days in V803 Cen, and result in the AM CVn achieving maximal brightness. Normal outbursts are far shorter events lasting between one and five days, with a peak brightness typically 1 magnitude dimmer than their superoutburst counterparts \citep{2012ApJ...747..117C}. SO generally have more complex profiles consisting of a sudden increase in brightness that is bluer in colour \citep[when compared to the system in quiescence;][]{2020A&A...636A...1H}, which gradually decreases over the duration of the outburst. Regardless of the duration of the SO, they often exhibit a small dip in brightness soon after maximum brightness after which the brightness can increase again \citep{2012MNRAS.419.2836R}.

Some systems also spend extended periods, sometimes several years, in ``high states'' similar to the ``standstills'' that are observed in Z Cam systems. Such states have been attributed to the mass accretion rate lying close to the critical value for accretion disc instability \citep{2001IBVS.5091....1K}. From a theoretical standpoint the behaviour of these systems is described by the disc instability model \citep[DIM;][]{1983A&A...121...29M}. This model outlines how different mass accretion rates, a key property driving AM CVn behaviour, and disc temperatures arise; which states are stable, and how the unstable systems can exhibit outbursts. Crucially DIM also predicts that the mass accretion rate is strongly correlated to the orbital period of a system, and thus is a marker for its expected behaviour \citep{2010PASP..122.1133S}.

KL Dra was one of the first AM CVn systems that was seen to exhibit regular outbursts \citep{2002MNRAS.334...87W}. Follow up work by \citet{10.1111/j.1365-2966.2010.17019.x} found that KL Dra showed SO approximately every 60 days, which lasted for about two weeks. Conversely, despite their almost identical orbital periods, \citet{2000MNRAS.315..140K} found that CR Boo appeared to show a cyclic behaviour moving between high and low states on a timescale of 46.3 days. In order to gain a better understanding of these differences we have selected a sample of AM CVn systems with orbital periods within 2.5 minutes of that measured for CR Boo and have collated photometric data from a range of All Sky Surveys and amateur measurements taken over a period of 15 years. This data allowed us to study the long term outbursting properties of these AM CVns with many having data for the entire 15 years considered -- although other systems such as CX361were observed less often. We further used \textit{TESS} full frame images to study KL Dra which revealed key detail in the SO for the first time.

\section{All Sky Surveys and AAVSO Photometry}
\begin{table}
\caption{Key observational parameters of the AM CVn systems in this study which have periods between 22.5-26.8 minutes. Periods taken from \citet{2018A&A...620A.141R}; all other data, including uncertainties, established in this work\label{AMCVN_table}. We show the orbital period, P\textsubscript{orb}, the SO recurrence time, T\textsubscript{Rec}, the SO amplitude, the SO duration, T\textsubscript{Dur} and not the presence or absence of normal outbursts.}
\begin{center}
\resizebox{\columnwidth}{!}{

\begin{tabular}{lccccc}
\hline
System&P\textsubscript{orb}&T\textsubscript{Rec}&Amplitude&T\textsubscript{Dur}&Normal\\
&(mins)&(days)&(mag)&(days)&Outbursts\\
\hline
PTF1 J1919+4815&22.5&35±1.4&2.3±0.1&17±0.4&\checkmark\\
ASASSN 14CC&22.5&29±7.1&3.9±0.1&14±2.2&\checkmark\\
CX361&22.9&-&-&-&$\times$\\
CR Boo&24.5&49±1.6&3.2±0.2&24±1.5&\checkmark\\
KL Dra&25.0&60±3.0&3.6±0.5&10±0.7&\checkmark\\
PTF1 J2219+3135&26.1&67.3±2.4&3.5±0.4&-&\checkmark\\
V803 Cen&26.6&74±1.4&4.2±0.5&28±3.8&\checkmark\\
PTF1 J0719+4858&26.8&-&2.9±0.1&-&$\times$\\
\hline
\end{tabular}}
\end{center}
\end{table}

\subsection{Data}

We combined data from a number of all sky surveys as well as data gathered by amateurs and made available through the AAVSO International Database \citep{AAVSO_CITE}. The surveys which we used were Catalina \citep{2009ApJ...696..870D}, ASAS-SN \citep{Kochanek_2017}, Pan-STARRS \citep{2020ApJS..251....7F}, ZTF \citep{2019PASP..131a8002B} and GOTO \citep[in preparation]{GOTO_proto} which brought together photometric data from 2005 through to 2020. Data from Catalina, Pan-STARRS, ASAS-SN and ZTF was accessed through their respective public data releases. Additional, more recent, ZTF data was also acquired through the Lasair transient broker \citep{2019RNAAS...3...26S}. Data from GOTO was accessed through the in-collaboration database which offers access to photometry shortly after observation.

\begin{figure*}
\includegraphics[width=\textwidth]{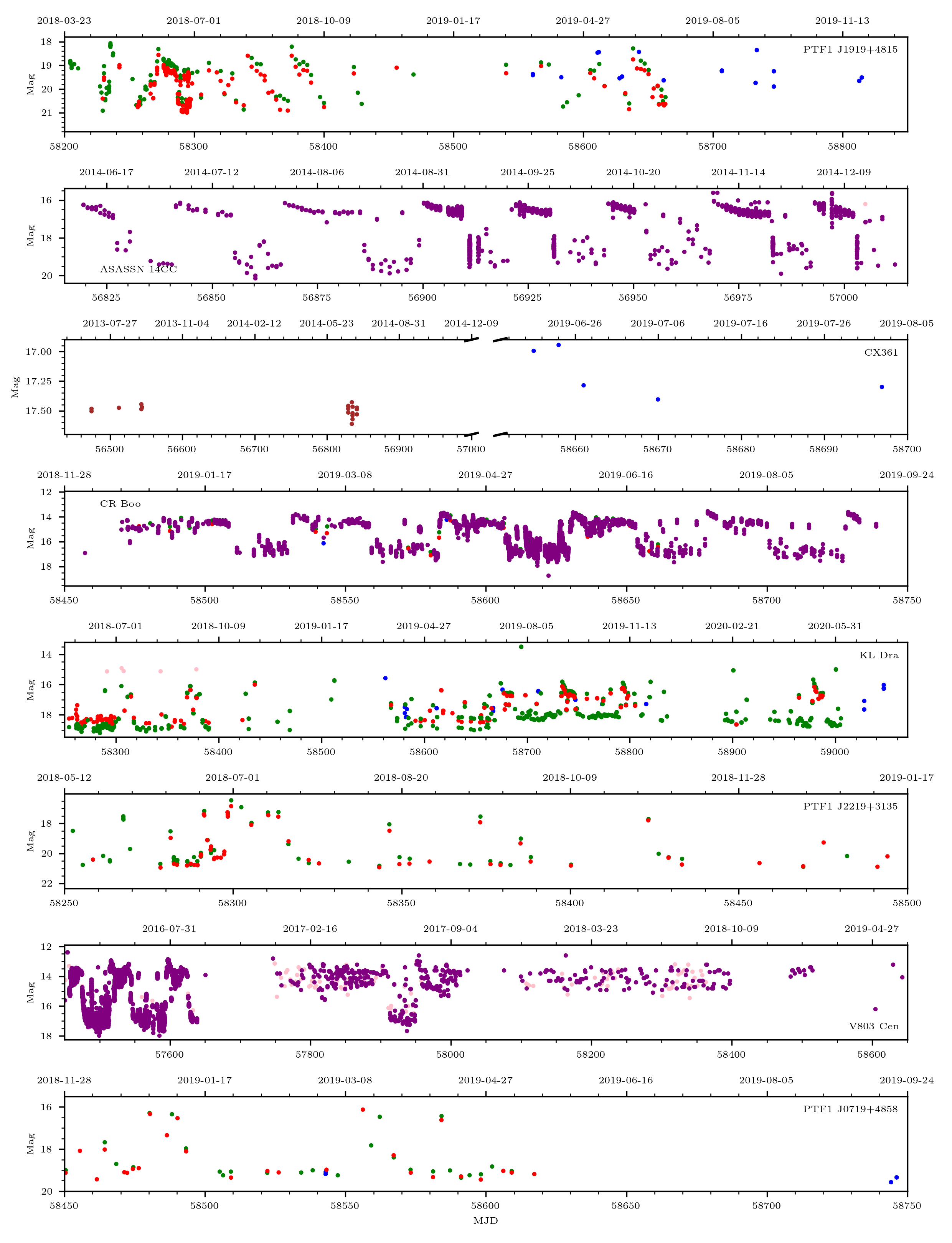}
\caption{Representative subsections of each of the light curves of the sources under investigation. Blue points from GOTO, red and green points from g and r band ZTF data, pink points from ASASSN, brown points from Pan-STARRS and purple points from AAVSO.}
\label{fig:repcurves}
\end{figure*}

As the data we used came from a range of telescopes we had to take care to select data from suitable band passes to ensure compatibility. For Pan-STARRS, ASAS-SN, and ZTF this was achieved using data taken in their respective \textit{g} band ($\sim4087-5522$\AA) filters. In ZTF we also accessed \textit{r} band data which is used illustratively in some figures but was not part of the analysis. Similarly for GOTO the data used was taken in either the \textit{L} band ($\sim4000-6800$\AA), or with a clear filter. Data obtained from Catalina were limited to the \textit{V} band which has significant overlap with these other filters ($\sim 4740-6860$\AA). Similar consideration was made to the AAVSO data and only data acquired using a \textit{V} band filter or no filter but calibrated with a \textit{V} band zero point was used. \autoref{fig:repcurves} shows subsections of the light curves of each of the sources which we have studied in this paper combining each of these data sets.

In surveys with quality flags associated with each photometric measurement, such as in Pan-STARRS, these were used to identify bad data which should not be included. Manual filtering, by e.g. inspecting source images, was also performed with measurements which were considered to be the most flawed also removed, e.g. those subject to poor seeing. This conservative approach, usually removing fewer than 5 data points, ensured that measurements which could be part of a real feature are not inadvertently removed whilst also avoiding introducing data that maybe indicative of a feature that does not exist.

\subsection{Data Analysis}\label{data_ana}

In order to search for any periodicity in systems, such as regular SO, an Analysis of Variance (AOV) period search \citep{1989MNRAS.241..153S,2005ApJ...628..411D} from the \texttt{vartools} suite developed by \citet{2016A&C....17....1H} was employed. AOV, which is well suited to determining the orbital periods of eclipsing binaries, works on the principle of folding and binning the data on different periods and identifying the period which minimises the difference between data points in the same bin from successive cycles. The results of these period searches, from successive observing seasons, provided our recurrence times for SO. In addition to this, since there were differences in the sampling frequency of observations, visual inspection was used to verify results. The \texttt{Lightkurve} \citep{2018ascl.soft12013L} python module was used to fold the photometric data according to the identified period. Doing this made it possible not only to see the overall cyclic behaviour in an observing season but it also allowed for cycles to be compared as per \autoref{fig:CRBoo2019} and \autoref{fig:CRBoo2019Cycles}. Consequently it was possible to develop an understanding of the differences in those features  which were present or absent in different cycles and observing seasons. 

Using plots such as these we were able to study the SO cycles of the systems in detail visually identifying the duration and amplitude of SO, the existence of normal outbursts and any dips in SO brightness. \autoref{AMCVN_table} shows the key properties determined for each system in our study; the mean recurrence time, amplitude, SO duration and the presence of normal outbursts, the errors on these mean values were calculated from their respective standard deviations. This high level comparison shows that despite all of these systems having relatively similar orbital periods, the property considered to be key in determining their behaviour, they appear to show quite divergent behaviour.

What follows is a discussion of the observed behaviour of those systems with periods within 2.5 minutes of CR Boo as seen in our data. Note that when individual observing seasons are discussed in this paper they are referred to by the year which forms the majority of the observing seasons. 

\subsubsection{PTF1 J1919+4815}
The behaviour that is exhibited by PTF1 J1919+4815 shows two distinct states which it cycles between. The SO which marks this transition lasts on average 17±0.4 days and occurs on average every 35±1.4 days; meaning that the system spends an equal amount of time in high and low states. Our results are consistent with a previous study of this system which identified a recurrence time of 36.8 days with a duration of $\sim13$ days \citep{2014ApJ...785..114L}. From the data available for this systems it appears that this is a consistent behaviour seen in all observing seasons considered. The SO show a minor dip and then increase in brightness a few days after onset. The low state for this system is close to the limiting magnitude for the surveys we used; however as we see a number of short lived events of magnitude less than that seen in SO we are confident that the system does exhibit normal outbursts, which have also been observed in previous studies.

\subsubsection{ASASSN 14CC}
ASASSN 14CC was discovered in 2014 by ASAS-SN \citep{2019MNRAS.486.1907J} and was subsequently observed extensively by amateur astronomers in the proceeding months; consequently there is a wealth of data from 2014, however following this, amateur observation appeared to cease leaving only limited data from All Sky Surveys. Nevertheless the data from 2014 is sufficient to allow discussion of the observed behaviour. ASASSN 14CC clearly shows distinct high and low states with abrupt changes from one to the other. The 2014 observations show 7 distinct high states with 7 corresponding low states. ASASSN 14CC appears to spend the majority of its time in a high state with these lasting for a mean duration of 29±7.1 days; this contrasts with low states which have a relatively constant 14 day duration. As is often seen in AM CVn systems the low states of ASASSN 14CC show frequent normal outbursts. A higher density of long term observations than that available to us would be required to confirm the continuity of this behaviour but the limited data we had access to suggested that this was archetypal of the system. The light curve of ASASSN 14CC is reminiscent of that of CR Boo (see \textsection\ref{CRBOO}). If this is the case then we would expect ASSASN 14CC to show extended periods of time in only a high state; we do not see this in our data, but encourage future observations at high cadence to establish if ASASSN 14CC shows this behaviour.

\subsubsection{CX 361}\label{sec:361}
First discovered by \citet{10.1093/mnrasl/slw141}, CX 361 was identified as an AM CVn with an average apparent \textit{i} band magnitude of 17.35 with short term variability of 0.2 mag over a period of 15 years as seen in the OGLE III and IV surveys. In addition to this a long term modulation, where the brightness appears to trend downwards was also observed. During the 15 years of observations it was not seen to exhibit any outbursts and was further identified as being in a high state from optical spectroscopic observations. This marks the system as highly unusual as it lies outwith the previously defined zones of stability for accretion discs and would be expected to outburst regularly.

In the data which we considered here CX 361 has only been observed by GOTO and Pan-STARRS each providing a small number of observations. The crowded nature of the field (CX 361 is located in the galactic bulge) makes extracting photometry challenging; however our data appears to be consistent with the earlier observations of \citet{10.1093/mnrasl/slw141}. Although we have only limited data we see some evidence, particularly in Pan-STARRS, that the previously identified gradual decrease in brightness has continued. Much of the differences which we see in our data may be attributed the use a different filter and observations crowded fields being highly sensitive to the effects of variable seeing conditions making it harder to resolve single sources.

\subsubsection{CR Boo}\label{CRBOO} 

\begin{figure}
\includegraphics[width=\columnwidth]{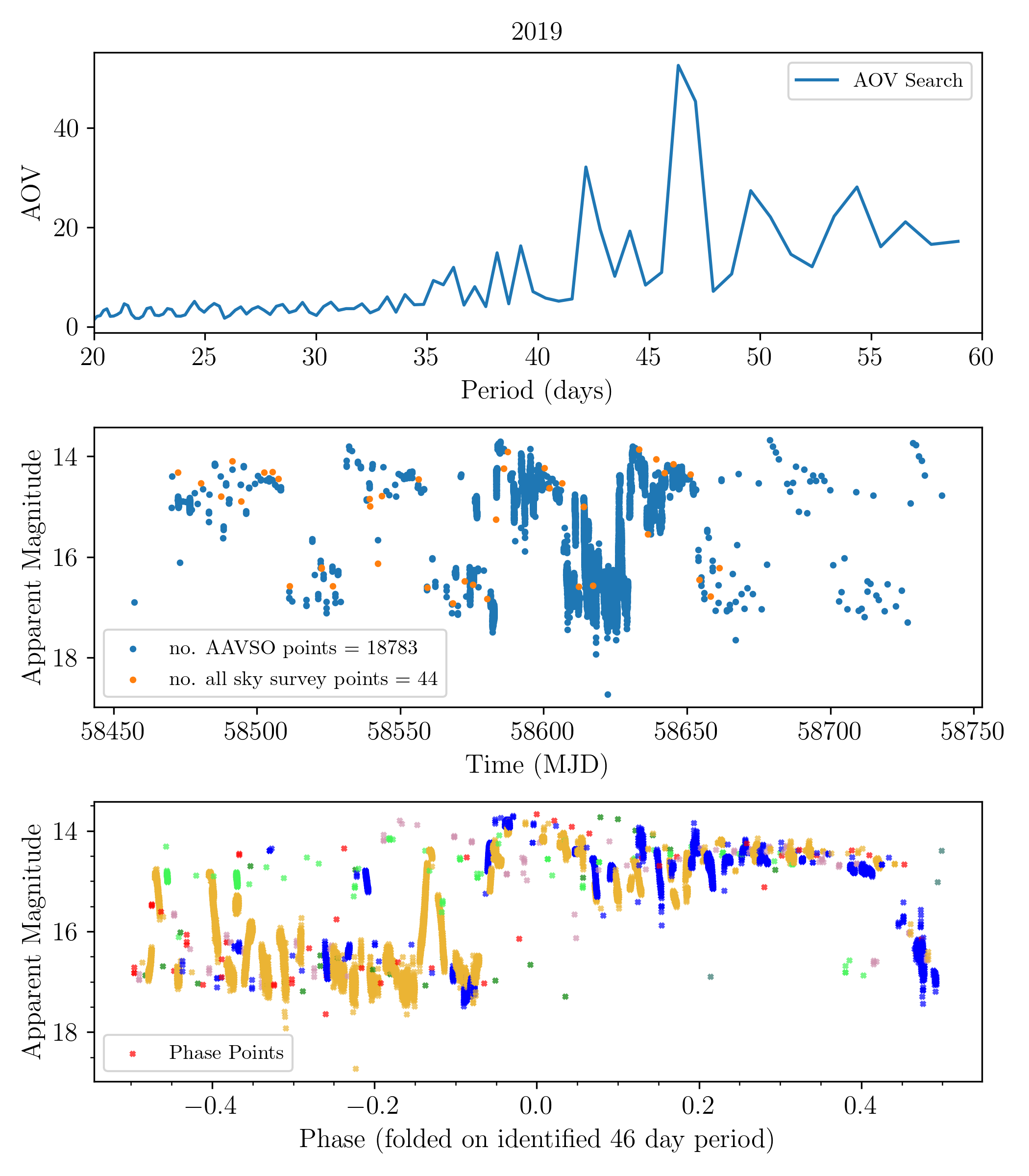}
\caption{Upper: AOV periodogram of the CR Boo 2019 observing season. Middle: light curve of CR Boo in the 2019 observing season. Lower: phase folded light curve of the 2019 observing seasons. The different colours represent a different cycle and phase 0 corresponds to the mean transition time to the high state for each of the cycles. We see a remarkable consistency from cycle to cycle in the high state behaviour, in addition to numerous normal outbursts during the low state.}
\label{fig:CRBoo2019}
\end{figure}

\begin{figure}
\includegraphics[width=\columnwidth]{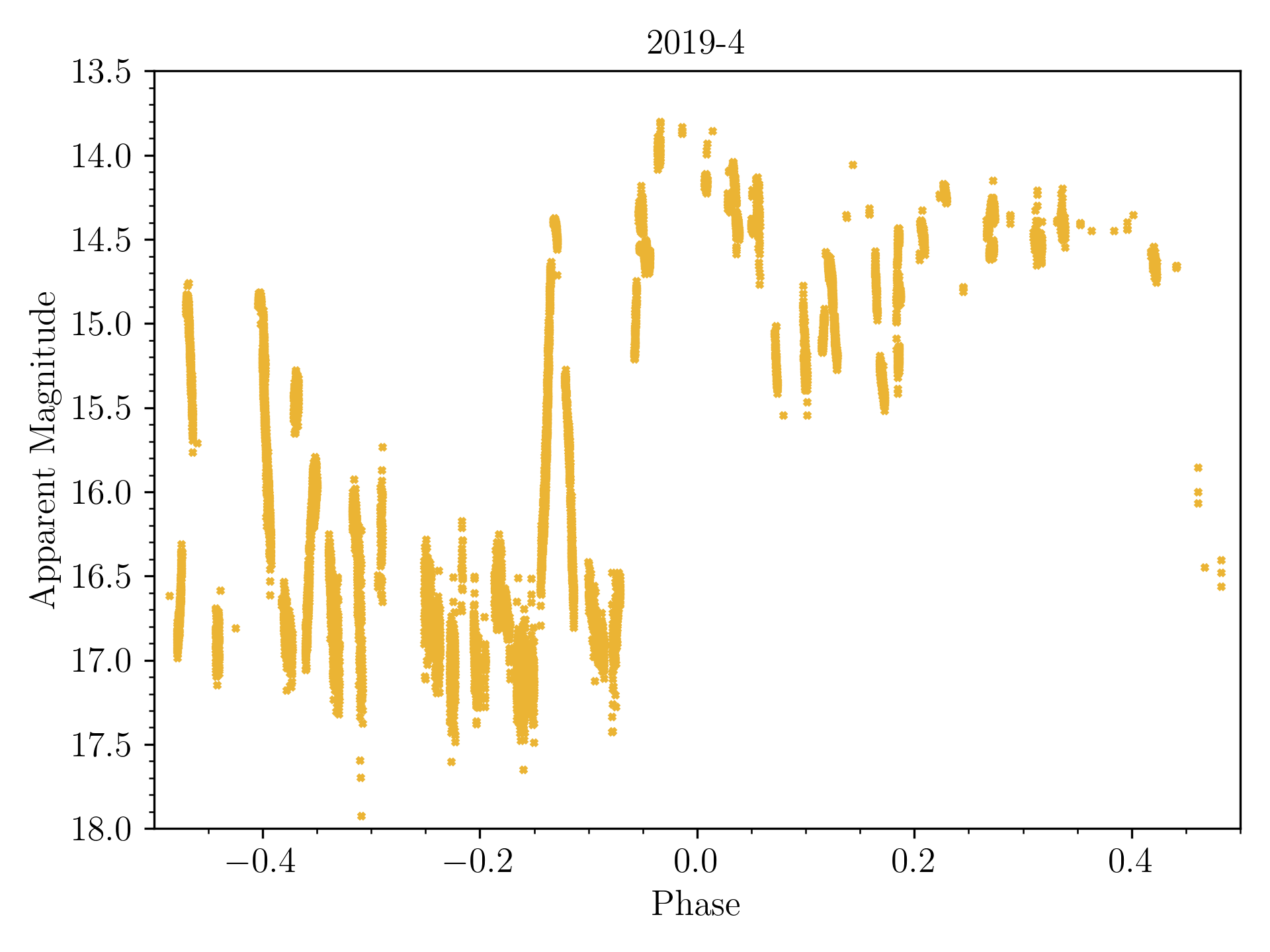}
\caption{The light curve of the CR Boo SO cycle 2019-4 extracted using the inferred period as shown in \autoref{fig:CRBoo2019} (48.5days).}
\label{fig:CRBoo2019Cycles}
\end{figure}

Plots such as those shown in \autoref{fig:CRBoo2019} and \autoref{fig:CRBoo2019Cycles} allowed us to extract the period of the SO cycle and study individual cycles for different observing seasons of CR Boo. Using these plots it was possible to make a detailed study of observing seasons from 2005 through to 2019. From this study it was clear that there are two types of seasonal behaviour; those which only show a continuous high state and those which alternate between a high and a low state. Those seasons which only show a high state make up $\sim40\%$ of the seasons studied and show an absence of other features in their light curves although occasionally they show short-lived dimming events. All but one of these seasons are seen to occur successively with other seasons of this type, indeed the early time of the 2018 season also exhibits this behaviour. The extended duration that this state is maintained for suggests that it is a stable configuration for CR Boo's accretion disc.

Those seasons which show both high and low states are far more feature rich in comparison. In those seasons that do exhibit a SO, e.g. at MJD = 58640, we see this to occur on a period that is centred upon 48 days. In those seasons which show low states, we see frequent (every 6 to 10 days) and sudden brightening events which we identify as normal outbursts similar to those seen by \citet{2012MNRAS.419.2836R}. We expect that these probably occur in every low state but due to their brevity have escaped observation in some low states as a result of the sampling. These observations compare favourably to those of \citet{2000MNRAS.315..140K} who saw the shift from high to low state occur on a period of 46.3 days and normal outbursts occur every 4-8 days. Within our observations, although broadly consistent, there is some drift season to season in observed SO recurrence time, with some seasons more in line with previous observations.

The other prominent feature frequently seen is a dip in brightness immediately after the onset of the SO i.e. the start of the high state, after which the brightness again increases. A clear example of this can be seen in cycle 2019-4 shown in \autoref{fig:CRBoo2019Cycles}. Although our observations do not show this feature in every season it is likely that the dip is a feature common to SO and that poor data collection, most likely poor sampling has resulted in the missed observation of it in some SO.

\subsubsection{KL Dra}
KL Dra has been observed by various All Sky Surveys which gave us access to several years worth of photometric data. It consistently exhibits SO which last between 7 and 15 days which recur on a timescale of approximately 60 days with an amplitude of 3.6 mag. This is consistent with the findings of \citet{2012MNRAS.419.2836R} who found a variable recurrence time centred on 60 days with durations in the same range that we observed. Similar to many of the other AM CVn systems in our study, the SO which we see also exhibit a dip in brightness shortly after onset. We discuss these SO in more detail using \textit{TESS} observations in \textsection\ref{tess_data}.

During the low states between successive SO we see 3-4 normal outbursts which appear to last approximately 1 day. Our observations of these normal outbursts are in line with the current understanding that KL Dra does exhibit normal outbursts, with previous work e.g. \citet{2012A&A...544A..13K} coming before this discovery and so building models without accounting for their existence.  

\subsubsection{PTF1 J2219+3135}
The behaviour observed in PTF1 J2219+3135 appears to be similar to that discussed above for KL Dra. Though data does not have particularly good sampling, a pattern similar to that seen in KL Dra can be identified. Using the AOV method we were able to extract a recurrence of time of approximately 67.3 days which we could tenuously confirm by visual inspection. Due to the sampling of the data it was not possible for us to determine an outburst duration, nor was it possible to confirm the presence of any dip in the SO. Furthermore our data does not show convincing evidence of normal outbursts in PTF1 J2219+3135.

\subsubsection{V803 Cen}
V803 Cen displays behaviour which is very similar to that seen in CR Boo; exhibiting both high and low states as well as extended periods where it is only found in a high state, which make up the majority of observations. As is common with other systems that exhibit low states, V803 Cen shows normal outbursts during the low states. The SO that V803 Cen exhibits also show a dip in their brightness immediately after their maximum brightness as is seen in other systems. The periodicity of the transition between the high and low states is somewhat hard to ascertain due to the apparent preference shown in many observing seasons towards the high state which limits the number of observed transitions. However in those seasons with transitions, the recurrence time, whilst varying by several days, lies at approximately 74 days. Of the SO that we did see, the majority had a duration, as determined by visual inspection of the lightcurves, centred on 28 days, although a few were seen to have durations as short as only 20 days. 

\subsubsection{PTF1 J0719+4858}
PTF1 J0719+4858 is one of the dimmest sources considered in this work regularly being observed at magnitude 20. As a result of this it is more sparsely observed that many of the other sources we have studied, nevertheless there is sufficient data to allow for a broad discussion of the behaviour observed. Although not densely sampled the system appears to show distinct high and low states which it transitions between; however with the data available it not possible to determine the timescale of these transitions, which state is preferred by the system, or if the SO contains any feature of note. Additionally the system does not appear to show any normal outbursts in the low states occurring the between the SO.

\section{TESS Observations}

\begin{figure*}
\includegraphics[width=\textwidth]{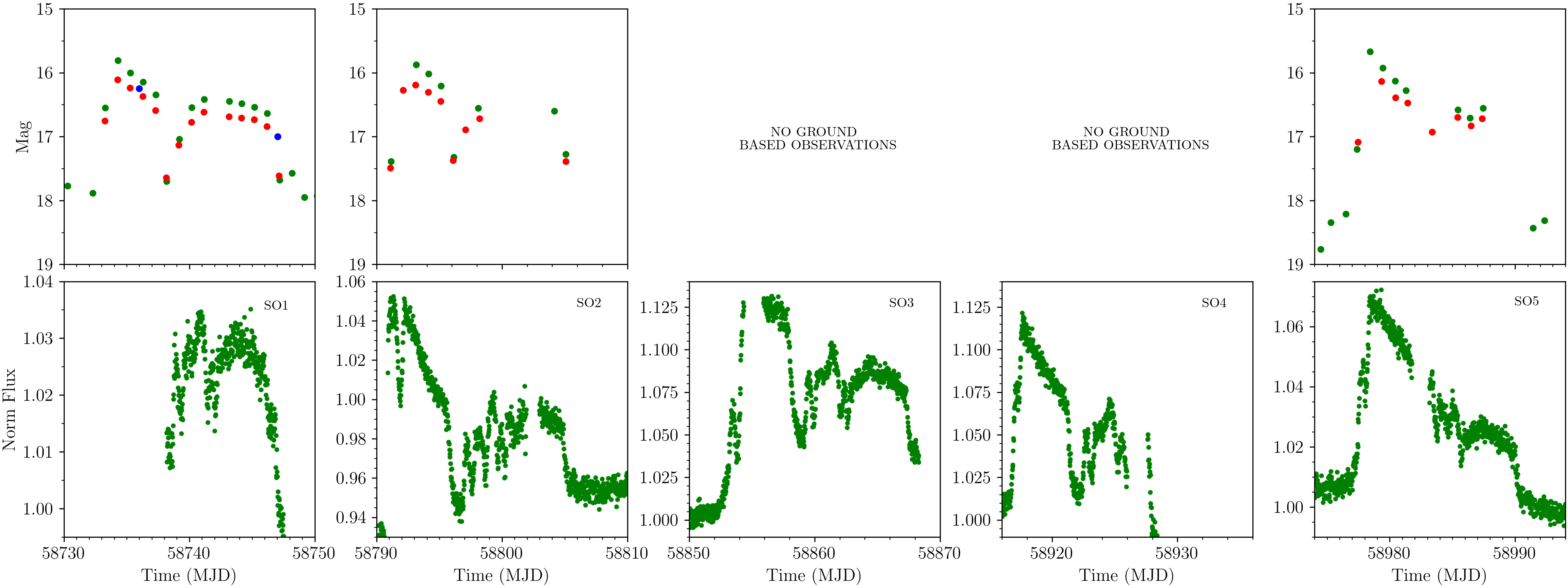}
\caption{Top panels: ZTF g and r band, in green and red, and GOTO, in blue, data showing SO in KL Dra. Bottom panels: TESS data taken contemporaneously showing each of these SO and the immediately proceeding normal outburst. No ground based data is available for SO3 and SO4 as KL Dra was too close to the Sun at these times.}
\label{fig:ztf-tess}
\end{figure*}

The \textit{Transiting Exoplanet Survey Satellite (TESS)} satellite was launched on 18$^{th}$ April 2018 into a 13.7 day orbit and has four small telescopes that cover a 24$\times$90\degr area of sky (see \citet{2015JATIS...1a4003R} for details). It has now covered both the southern and northern ecliptic hemispheres apart from a strip along the ecliptic plane. Each sector of sky is observed for around 28 days. In the first two years of observations, around 20,000 stars were observed with a cadence of 2 minute in each sector. However, the full frame images which we made use of only give a cadence of 30 minute for each sector. KL Dra is located in the continuous viewing zone near the northern ecliptic pole (data from sector 15 were not available). Ideally we would have gathered data of more than a single system, however CR Boo, the only other system which could have been seen in more than a single sector, repeatedly fell in a chip gap or just out of field. 

\subsection{Data Analysis}\label{tess_data}
One of the challenges of obtaining photometry of KL Dra using \textit{TESS} is the large pixel scale (21\arcsec/pixel) coupled with the fact that it is spatially nearby (6\arcsec) a galaxy \citep[it was originally misclassified as a supernova -- SN 1998di][]{1998IAUC.6983....1J,2002MNRAS.334...87W}. To obtain photometry we used packages \texttt{eleanor} (\citet{2019PASP..131i4502F} and \texttt{lightkurve} \citep{2018ascl.soft12013L} which correct for the background and remove instrumental effects which are present in the data. Both produced similar results but we present the photometry produced by \texttt{lightkurve}. We extracted 15$\times$15 pixel postage stamps centred on KL Dra. We took the flux from a 3$\times$3 set of pixels centred on KL Dra and took background from pixels which were below a low threshold.

We did not expect to detect KL Dra in quiescence since it is much fainter than the nearby galaxy. However, we hoped we may be able to detect both normal and SO in KL Dra. In \autoref{fig:ztf-tess} we show five time intervals where we have obtained contemporaneous photometry using ZTF and GOTO, although we do not have ground based data for two of the SO as KL Dra was too close to the Sun. Since ZTF data shows SO at the same time as found from \textit{TESS} observations, we are confident that \textit{TESS} has recorded five SO from KL Dra. This is the first time that any SO from a AM CVn has been observed at such high cadence over its duration.

In the first SO (SO1), observations started mid-way through the SO and there is a short gap near the peak of SO3. In four SO the dip which has been seen in previous observations of KL Dra and other AM CVn systems is also seen. After the flux has recovered from the dip, we find that in two SO, the source becomes highly variable. In all cases there is a rapid decline in flux at the end of the SO.

The difference in the apparent scale of the outbursts between the two sets of observations is immediately apparent, with the outburst in \textit{TESS} appearing far less prominent. Due to the large pixel size the nearby galaxy (which is significantly brighter than KL Dra) dominates the light in the pixels. We confirmed this effect by performing aperture photometry upon the sources in the approximate field of view of the \textit{TESS} pixel using data from Pan-STARRS and simulating the effect of a SO on the received flux. We found that a SO produces an approximate increase of 10\% in flux received which is in line with the \textit{TESS} observations.

\begin{figure}
\includegraphics[width=\columnwidth]{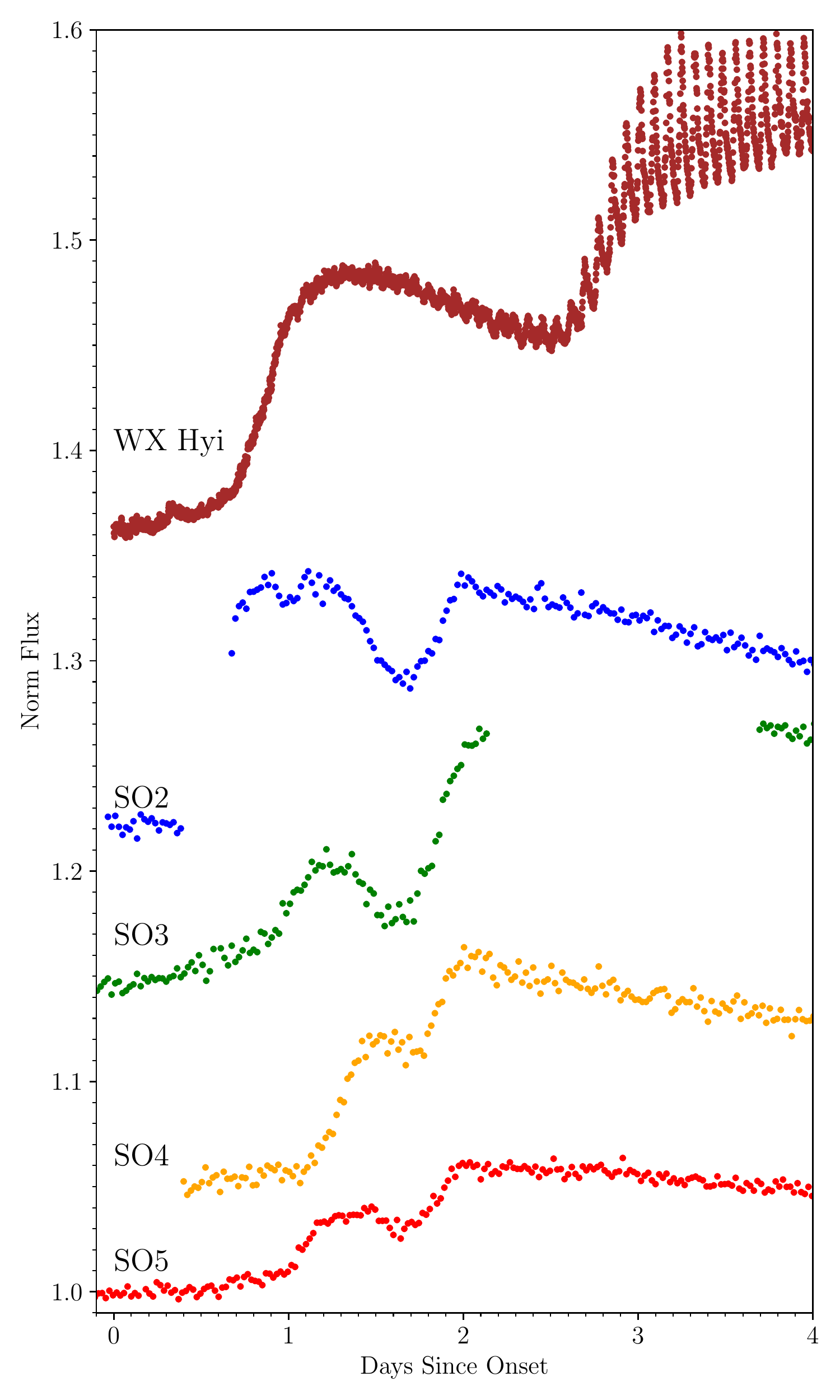}
\caption{Four SO from KL Dra and one from DN WX Hyi as seen by TESS. Each have been overlapped in time such that the dip following the proceeding outburst occurs at approximately 1 day after onset. The WX Hyi SO shows similar features although occurring on a longer time scale; we have compressed it for straightforward comparison. These observations of WX Hyi show evidence of superhumps, although previously observed in KL Dra these observations do not show this as a result of the flux excess from the other sources in the same pixel.}
\label{fig:tess-norm-sup}
\end{figure}

In \autoref{fig:tess-norm-sup} we show the first four days of four of the SO observed. We have manually shifted the time axis so the dip lines up. In each case we find a clear drop in flux around 1 day from the initial rise in flux. Similar behaviour was first reported in a SO of a hydrogen accreting dwarf novae using \textit{Kepler} data \citep{2012ApJ...747..117C}, where it was identified to be a normal outburst acting as the trigger of a subsequent SO. Due to the similarity, we attribute this finding in KL Dra to be the same feature. Further observations of other dwarf novae, show this to be a feature of all SO observed in high cadence. However, this is the first time that this precursor has been seen in an outburst from an AM CVn.

We also show one SO from the hydrogen rich accreting dwarf nova WX Hyi (using \textit{TESS} 2 minute data from sector 3). This shows a normal outburst preceding the SO: the difference is the timescale where the SO occurs over a longer timescale which is expected given the difference in scale of the binaries. These observations of WX Hyi show evidence of superhumps; although these have previously observed in KL Dra \citep{2002MNRAS.334...87W} they are not present in these TESS data. This, as with the apparent magnitude of the SO observed, is a direct result of the flux excess from the other sources lying in the same pixel.

We further identify a number of normal outbursts between the SO observed in KL Dra. Although there are short gaps in the data implying that some normal outbursts may have been missed, we find there are three or four normal outbursts between SO. Between SO1 and SO2 the time interval between normal outbursts is 6-10 days; SO2 and SO3 12-17 days; SO3-SO4 $\sim$12 days; SO4-SO5 $\sim$13 days and after SO5 $\sim$14 days. This appears to be correlated with the SO recurrence time with a shorter recurrence time leading to a shorter interval between normal outbursts. Observations such as these will inform future models of outbursts in systems similar to KL Dra. 

\section{Discussion}
We have presented optical photometry collected over 15 years of eight AM CVn systems all of which have an orbital period in the range 22.5 -- 26.8 min. The vast majority of these systems have been seen to outburst in a manner consistent with the findings of \citet{2012MNRAS.419.2836R} who identified that systems with an orbital period between 20 and 45 minutes undergo outbursts. The properties of these SO have been identified to be correlated with the period of the system as quantised by \citet[\textsection4.2]{2015MNRAS.446..391L} and \citet{2015ApJ...803...19C}.

These systems have now been sufficiently well sampled to allow us to distinguish between normal outbursts, which only last a few days at most, and SO, which last at least a week. This has allowed some of these systems to be identified as the hydrogen deficient analogues of SU UMa dwarf novae which show similar behaviour \citep{2012A&A...544A..13K}. Complete study of the normal outbursts has, however, only been possible with space telescopes such as TESS; offering sufficient temporal resolution to see these short lived events. Normal outbursts are seen clearly in PTF J1919+4815, ASASSN-14CC, CR Boo, KL Dra, PTF J2219+3135 and V803 Cen both in our observations and that of others. This has allowed for a more complete picture of the behaviour of these systems to be observed and thus address oversights in earlier modelling work which omitted features due to their apparent absence. 

\subsection{Outburst Properties and Period}

In our study of these AM CVn systems we identify two distinct groupings of outbursting systems (\autoref{fig:repcurves}). ASASSN 14CC, CR Boo and V803 Cen each show very similar high and low states of similar duration. Reminiscent of the standstills that are seen in Z Cam systems \citep{2014JAVSO..42..177S}, we see that CR Boo and V803 Cen have extended periods of time ($>1$ year), where they are only seen in a high state. The remaining systems, with the exception of CX 361, show regular but much shorter SO, and do not appear to show evidence of standstills.

\begin{figure}
\includegraphics[width=\columnwidth]{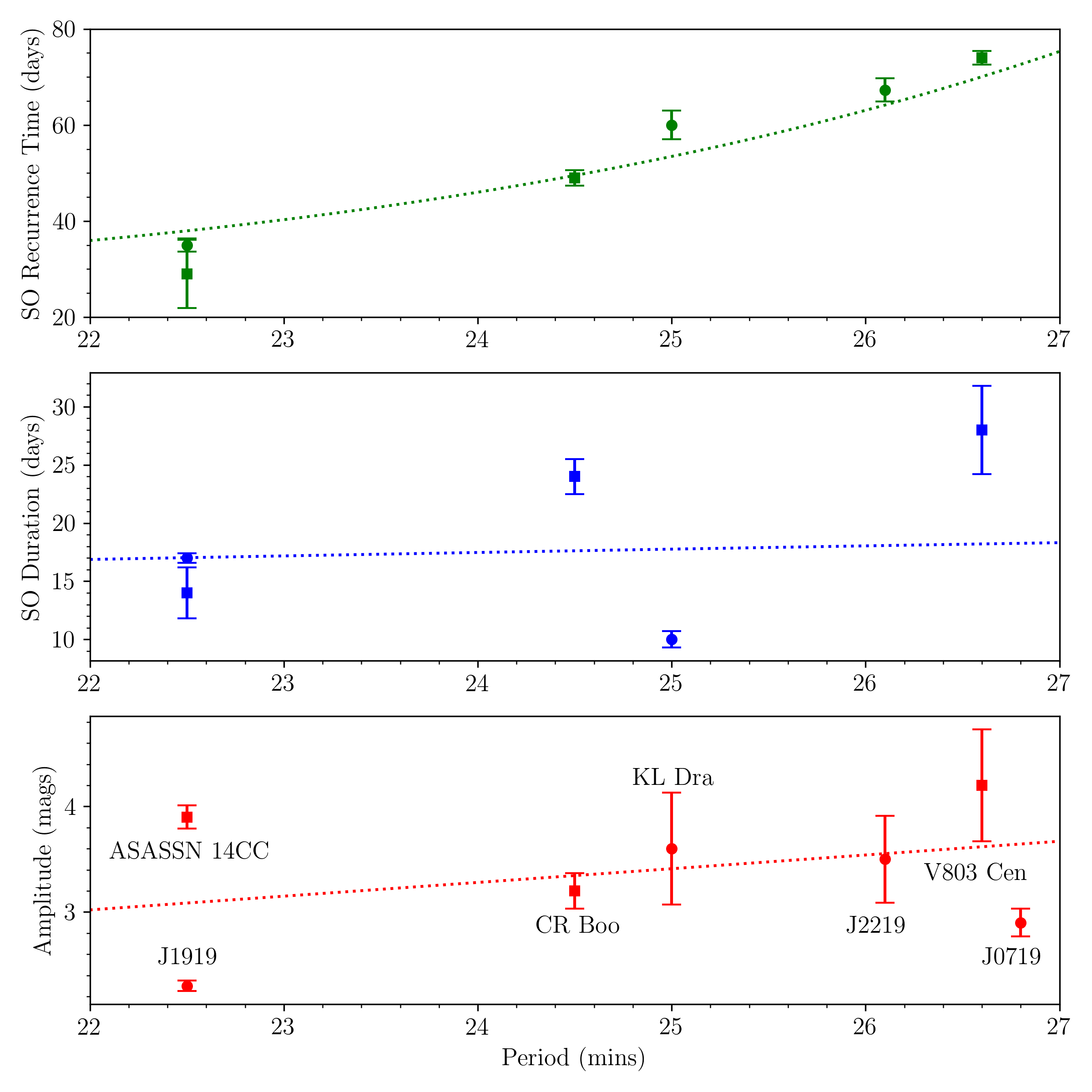}
\caption{From top to bottom, the SO recurrence time, duration, and amplitude plotting as function of orbital period. The lines represent the relationships for amplitude and recurrence time derived by \citet{2015MNRAS.446..391L}, and duration derived by \citet{2015ApJ...803...19C} as a function of orbital period. Squares and circles denote CR Boo-like and KL Dra-like systems respectively. Values taken from \autoref{AMCVN_table}, recurrence and duration for PTF1J0719+4858 from \citet[Table 4]{{2011ApJ...739...68L}}.}
\label{fig:functionOfPeriod}
\end{figure} 

In \autoref{fig:functionOfPeriod} we compare the properties of the SO from each of the systems which we have examined with the qualitative predictions made by \citet{2015MNRAS.446..391L} and the predictions based on the DIM made by \citet{2015ApJ...803...19C}, and refined by \citet{2019AJ....157..130C}. It is immediately clear that systems with very similar periods can have divergent outbursting properties. For example, our data indicates that in those observing seasons which show outbursting behaviour, CR Boo has a recurrence time that is centred on 48.6 days; conversely KL Dra, which has an orbital period only 30 seconds longer, shows a recurrence time on the order of 60 days. 

This is not the first time that such differences have been noted and various work has been undertaken to classify this behaviour as the analogue of different hydrogen dominated CVs (see \citealt[and references therein]{2012A&A...544A..13K}). Consequently several attempts to adapt pre-existing hydrogen dominated models have been undertaken in order to explain observations, however the question of why apparently similar AM CVn system behave so divergently has not been addressed conclusively.

In \autoref{tab:mass_ratio} we show the mass ratios for 4 systems arranged according to the outburst behaviour that they exhibit. We see that those systems which behave like CR Boo have consistently lower mass ratios i.e. the differences between their masses are more pronounced. It is our contention that this change in relative masses may be a significant component in determining mass transfer rate from the donor star and thus the outburst profile followed by any given outbursting AM CVn system. Inferring the mass ratio of an AM CVn system is a nontrivial task, which in many cases requires substantial observations of a system. For example the superhump excess which can give the mass ratio from empirical relations \citep{2006MNRAS.373..484K}: consequently, mass ratio values are unavailable for a number of systems. Despite this, it provides a possible marker as to the origin of the divergence in outbursting AM CVn behaviour.

\begin{table}
\caption{Superhump derived mass ratios from \citet{2018MNRAS.477.5646G}. The left hand columns show those systems which we have identified to have long SO and standstills, whilst the right most columns show CX 361, a high state system, and and KL Dra which shows short SO. \label{tab:mass_ratio}}
\begin{center}
\begin{tabular}{cc|cc}
\hline
System&\textit{q}&System&\textit{q}\\
\hline
CR Boo & 0.058±0.008& CX361 & 0.070± 0.007\\
V803 Cen &0.058±0.014  & KL Dra & 0.092±0.006\\
\hline
\end{tabular}
\end{center}
\end{table}

There are however other possible factors which could be involved in determining the observed behaviour. The rate of mass transfer from the donor star is the fundamental property which determines outburst behaviours and there are a number of different ways in which this can manifest differently. The type of the donor star and the formation channel that was followed by the AM CVn system at its birth can affect this mass transfer rate, and incidentally affect the value of \textit{q} for a system. The mass transfer rate can also be affected the entropy of the donor and how it has been effected by irradiation \citep{2007MNRAS.381..525D}. These factors are not readily determined through observation alone so future confirmation of this will rely heavily upon modelling. An additional factor which could affect the accretion process is if the primary white dwarf had a sufficiently high magnetic field to either truncate or prevent the formation of an accretion disc. One AM CVn system (SDSS J0804+1616) has some evidence for the white dwarf having a significant magnetic field \citep{2009MNRAS.394..367R}.

In addition to the two different outbursting states we also see evidence of high state systems in our period range. In \textsection\ref{sec:361} we discussed the behaviour which we see in our data of CX 361. Consistent with previous observations, this system is unexpectedly deviates from theoretical predictions and is observed in a high state. Recently \citet{2020arXiv200902567B} identified another high state AM CVn system, ZTF J2228+4949, with a period of 28.6 minutes. Although this system lies outwith the period range we have studied, it does lie well within the range of periods that are predicted to outburst. Comparison of the spectra of CX361 and J2228+4949 with the high state spectra of KL Dra \citep{10.1111/j.1365-2966.2010.17019.x} yields no immediate evidence of difference. This is in line with expectations as they were identified as high state systems initially from their spectra, nevertheless it is possible that unusual metallicity is a component. Alternatively it is equally possible, as before, that another factor, such as a different formation channel or donor type has resulted in this permanent high state.

In high state systems, such as AM CVn itself, the state is maintained because the accretion disc temperature is always in excess of the ionisation temperature of Helium. This makes the temperature sufficiently high so as to have a mass accretion rate above the critical value; this in turn is a stable thermal equilibrium for the system to occupy and thus the system remains in a high state \citep[see \textsection 3.5.3.2 and 3.5.3.3]{warner2003cataclysmic}. 

The existence of such systems in the instability region is further proof that more than simply the orbital period of system is important when considering their behaviour. These systems, and other subsequently identified, should be a focus of further observations in order to identify which physical parameters are important in determining the behaviour of these systems and how the disc accretion model can be altered in light of this.

In previous studies (eg. \citealt{2012MNRAS.419.2836R,2015MNRAS.446..391L}), considering the AM CVn systems discovered at the time, it was possible to say that all systems within the period range $\sim 22$ and 44 minutes showed outbursts, however with discovery of further systems this is clearly not the case. Systems within this period range cannot be treated like an homogeneous group and like their hydrogen dominated cousins they ought to be subdivided based on their outburst behaviour. The high state systems appear to be the analogues of nova-like systems. Likewise, we agree with \citet{2012A&A...544A..13K} that KL Dra-like systems are most likely the helium dominated analogues of SU UMa systems whilst CR Boo-like systems are Z Cam analogues. In Z Cam systems the mass accretion rate lies very close to the critical mass accretion rate and standstills are believed to occur when this critical value is reached. it is equally possible however, that the CR Boo systems are analogous to VY Scl systems which show similar extended high states. It is still an open question as to what causes these variations in mass accretion rate but star spots have been suggested as a possible origin \citep{1994ApJ...427..956L} and it could be the case that systems of this type are more likely to host star spots.

\subsection{Precursor Normal Outbursts} \label{sec:Precursor}
Observations of dwarf nova using \textit{Kepler} \citep{2012ApJ...747..117C} provided evidence of normal outbursts immediately before the onset of a SO. Subsequent observations of SS Cyg using AAVSO data by \citet{2012ApJ...757..174C} also found evidence of such precursor outbursts. These were identified to be associated with the superhump where the mass transfer rate undergoes a modulation allowing for a brief cooling front to propagate.  

Since these discoveries it has been established that such precursor outbursts are seen in all dwarf nova. Until now these had not been seen in hydrogen deficient systems. Our observations of KL Dra in TESS provide the first evidence of a normal outburst immediately proceeding a SO in an AM CVn. The relative rarity of AM CVn systems has meant that it has only been possible to make this discovery with high cadence observations.

This has significant implications for the application of the disc instability model as applied to AM CVn systems; in dwarf nova they are believed to be the event that induces the subsequent outburst, their presence here suggests a similar effect. We believe that a similar superhump feature should be introduced to modelling of AM CVn SO to account for this discovery.

In order to identify if this is a feature common to all outbursting AM CVn systems, or indeed a subset of them, high cadence photometry of similar type to that supplied by TESS should be employed in future studies of these systems if possible. We believe, however, that evidence from hydrogen dominated systems points to this being likely.

\subsection{Dips}
Most of the sources in this study appear to show a pronounced dip in brightness a handful of days after maximum brightness which lasts for approximately a day. This feature has has previously been observed by \citet[\textsection4]{2012MNRAS.419.2836R} in various AM CVn systems. In our observations we see this dip feature, to varying degrees, in PTF1919+4815, ASASSN 14CC, CR Boo, KL Dra, and V803 Cen. Examples of these features can be seen in \autoref{fig:CRBoo2019Cycles} and \autoref{fig:ztf-tess}. For those systems where we do not see a dip, there is limited data, and previous work has seen a dip in PTF1 J0719+4858 \citep{2011ApJ...739...68L}, suggesting they are common in outbursts from AM CVn binaries.

\begin{figure}
\includegraphics[width=\columnwidth]{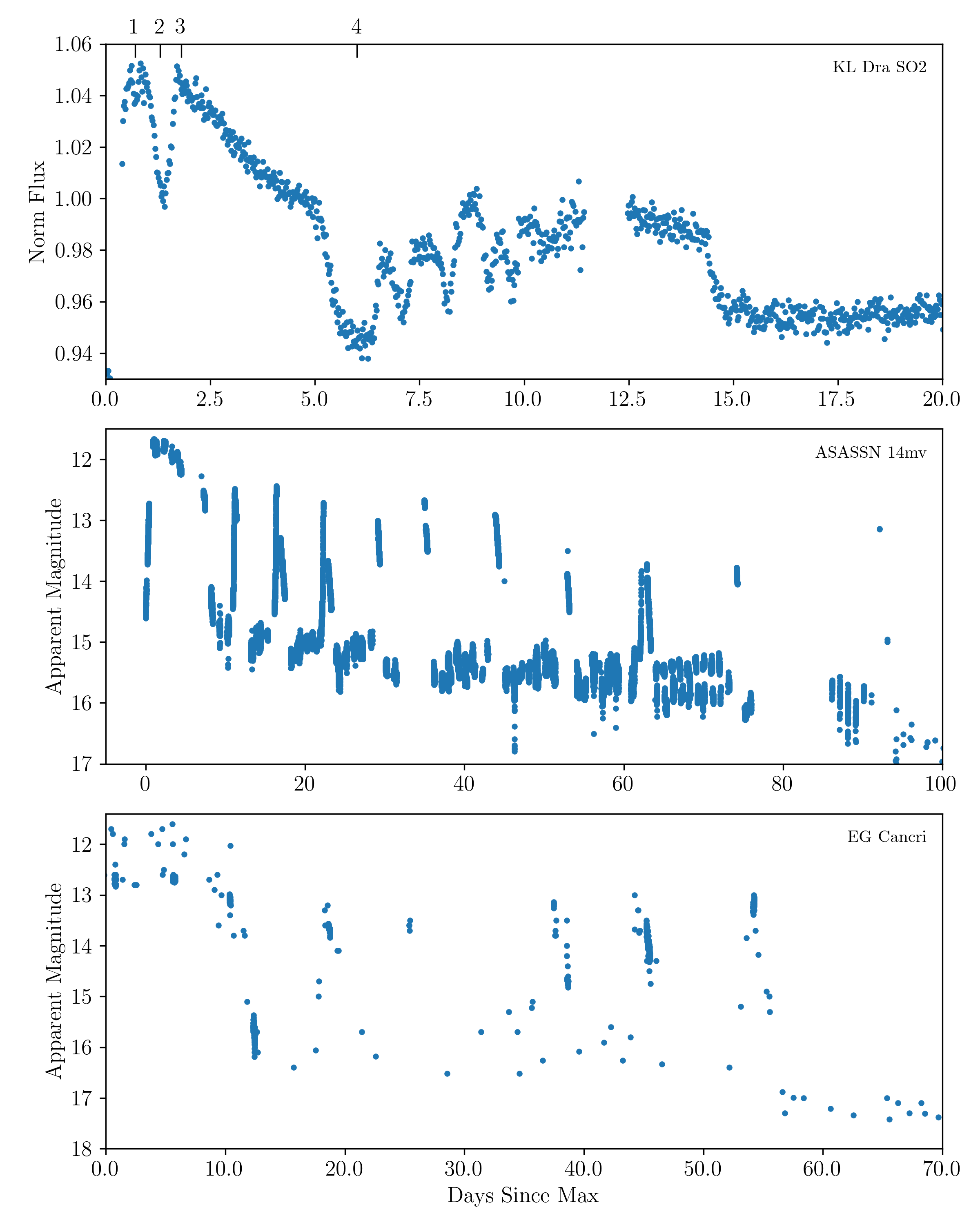}
\caption{Lightcurves of a SO observed in KL Dra using TESS and of ASASSN 14mv and EG Cnc, where data originates from AAVSO. In KL Dra \textit{1} denotes the peak of the preceding normal outburst, \textit{2}, denotes the dip in brightness between the normal outburst and the SO (\textsection\ref{sec:Precursor}), \textit{3} denotes the peak of the SO and \textit{4} denotes the dip in the SO. The middle panel shows an echo outburst in 2015 from the AM CVn binary ASASSN 14mv and the lower panel shows an echo outburst from the hydrogen accreting binary EG Cnc from 1996. Such features should inform models which predict outburst from accreting binaries.} 
\label{fig:dips}
\end{figure}

In \autoref{fig:dips} we compare one SO from KL Dra observed using TESS with two other systems: ASASSN 14-mv, a 41 minute AM CVn system \citep{2018A&A...620A.141R}, and EG Cnc, a SU UMa dwarf nova \citep{1998PASP..110.1290P}, both of which were identified as showing ``echo" outbursts, where multiple, shorter duration, lower amplitude outbursts are seen during the decline to quiescence. Although we expect the physical mechanism which gives rise to the dips seen in some AM CVn systems and the echo outbursts will be different, they highlight the need to obtain photometry covering the duration of the SO so that they can inform and test models which predict the properties of the accretion disc during an outburst (e.g. \citet{2012A&A...544A..13K}).

\section{Conclusions}
We have investigated the behaviour of 8 AM CVn systems in the disc instability region with orbital periods between 22.5 and 26.8 minutes in order to probe the behaviour of these systems and their outbursts. We present the first evidence of precursor normal outbursts before the onset of the SO in KL Dra; which we believe, based on hydrogen dwarf nova, are likely to be ubiquitous in AM CVn systems. If confirmed, this finding should prompt changes to the existing models for disc accretion in AM CVn systems.

We have further identified that the previously broadly accepted finding that systems with orbital periods between 22 and 44 minutes should exhibit outbursts is not as well defined as previously thought. We have studied one AM CVn and discussed another with periods inside this range that have only been seen to exist in their high state. It is also likely that more such systems exist but observational bias means that they have not been identified; outbursting sources often make more attractive targets for study meaning that these systems may well have been neglected. This adds to the growing body of evidence that suggests that the long term behaviour of AM CVn systems in the instability region is more subtle than previously thought and mirrors that of the hydrogen dominated accreting dwarf novae. We believe that further observations of more sources will cement this conclusion, yielding further evidence of these high state systems.

\section*{Acknowledgements}
The Gravitational-wave Optical Transient Observer (GOTO) project acknowledges the support of the Monash-Warwick Alliance; University of Warwick; Monash University; University of Sheffield; University of Leicester; Armagh Observatory \& Planetarium; the National Astronomical Research Institute of Thailand (NARIT); Instituto de Astrof\'{i}sica de Canarias (IAC); University of Portsmouth; University of Turku, and the UK Science and Technology Facilities Council (STFC grant numbers ST/T007184/1, ST/T003103/1).

We acknowledge with thanks the variable star observations from the AAVSO International Database contributed by observers worldwide and used in this research.

Armagh Observatory \& Planetarium is core funded by the Northern Ireland Executive through the Department for Communities. C. Duffy acknowledges STFC for the receipt of a postgraduate studentship.

D.M.S and M.R.K acknowledge support from the ERC under the European Union’s Horizon 2020 research and innovation programme (grant agreement no. 715051; Spiders).

The CSS survey is funded by the National Aeronautics and Space Administration under Grant No. NNG05GF22G issued through the Science Mission Directorate Near-Earth Objects Observations Program.  The CRTS survey is supported by the U.S.~National Science Foundation under grants AST-0909182.

The Pan-STARRS1 Surveys (PS1) and the PS1 public science archive have been made possible through contributions by the Institute for Astronomy, the University of Hawaii, the Pan-STARRS Project Office, the Max-Planck Society and its participating institutes, the Max Planck Institute for Astronomy, Heidelberg and the Max Planck Institute for Extraterrestrial Physics, Garching, The Johns Hopkins University, Durham University, the University of Edinburgh, the Queen's University Belfast, the Harvard-Smithsonian Center for Astrophysics, the Las Cumbres Observatory Global Telescope Network Incorporated, the National Central University of Taiwan, the Space Telescope Science Institute, the National Aeronautics and Space Administration under Grant No. NNX08AR22G issued through the Planetary Science Division of the NASA Science Mission Directorate, the National Science Foundation Grant No. AST-1238877, the University of Maryland, Eotvos Lorand University (ELTE), the Los Alamos National Laboratory, and the Gordon and Betty Moore Foundation.

Based on observations obtained with the Samuel Oschin 48-inch Telescope at the Palomar Observatory as part of the Zwicky Transient Facility project. ZTF is supported by the National Science Foundation under Grant No. AST-1440341 and a collaboration including Caltech, IPAC, the Weizmann Institute for Science, the Oskar Klein Center at Stockholm University, the University of Maryland, the University of Washington, Deutsches Elektronen-Synchrotron and Humboldt University, Los Alamos National Laboratories, the TANGO Consortium of Taiwan, the University of Wisconsin at Milwaukee, and Lawrence Berkeley National Laboratories. Operations are conducted by COO, IPAC, and UW.

\section*{Data Availability}
The following data used in this article is available in the public domain at the following locations Pan-STARRS: \url{https://catalogs.mast.stsci.edu/panstarrs/}, ASAS-SN: \url{https://asas-sn.osu.edu/photometry}, Catalina:  \url{http://nesssi.cacr.caltech.edu/DataRelease/}, AAVSO: \url{https://www.aavso.org/data-download}, and ZTF: \url{https://irsa.ipac.caltech.edu/Missions/ztf.html} and \url{https://lasair.roe.ac.uk}.

GOTO data products will be available as part of planned GOTO public data releases.

\bibliographystyle{mnras}
\bibliography{References.bib}

\bsp	
\label{lastpage}
\end{document}